\documentclass[conference,9pt]{IEEEtran}
\IEEEoverridecommandlockouts

\usepackage{cite}
\usepackage{amsmath,amssymb,amsfonts}
\usepackage{algorithmic}
\usepackage{graphicx}
\usepackage{textcomp}
\usepackage{xcolor}

\def\BibTeX{{\rm B\kern-.05em{\sc i\kern-.025em b}\kern-.08em
    T\kern-.1667em\lower.7ex\hbox{E}\kern-.125emX}}
\begin{document}

\title{SCDiar: a streaming diarization system based on speaker change detection and speech recognition}

\author{\IEEEauthorblockN{Naijun Zheng, Xucheng Wan, Kai Liu and Zhou Huan}
\IEEEauthorblockA{IT Innovation and Research Center, Huawei Technologies, China\\
Email: \{zhengnaijun, wanxucheng, liukai89, zhou.huan\}@huawei.com}
}

\maketitle

\begin{abstract}
In hours-long meeting scenarios, real-time speech stream often struggles with achieving accurate speaker diarization, commonly leading to speaker identification and speaker count errors.  
To address this challenge, we propose SCDiar, a system that operates on speech segments, split at the token level by a speaker change detection (SCD) module. Building on these segments, we introduce several enhancements to efficiently select the best available segment for each speaker. These improvements lead to significant gains across various benchmarks. Notably, on real-world meeting data involving more than ten participants, SCDiar outperforms previous systems by up to 53.6\% in accuracy, substantially narrowing the performance gap between online and offline systems.
\end{abstract}

\begin{IEEEkeywords}
diarization, speaker change detection, token-level, speaker-attributed ASR.
\end{IEEEkeywords}

\section{Introduction}
\label{sec:intro}
The streaming speaker diarization (SD) technologies are well-known to solve 'who spoke when' issue given an audio stream input. General SD systems can be broadly be divided into two categories: cascaded systems and end-to-end systems, differentiated by how they estimate the number of speakers. Cascaded systems use unsupervised clustering algorithms, such as spectral clustering \cite{Ning2006ASC} and VBx \cite{landini2022bayesian}. While the end-to-end systems utilize neural networks trained directly with diarization loss, such as End-to-End Neural Diarization (EEND) \cite{Fujita2019EndtoEndNS, Horiguchi2020EndtoEndSD} and Target-Speaker Voice Activity Detection \cite{Medennikov2020TargetSpeakerVA}.
Both types of systems have been adapted for online meeting scenarios, such as core sample selection \cite{Yue2022OnlineSD} with applying VBx and frame-wise streaming EEND \cite{Liang2023FrameWiseSE}. Additionally, approaches that combine unsupervised clustering and neural networks have been proposed \cite{Horiguchi2022OnlineND} to handle prolonged meetings.

In real meeting scenarios, simply knowing 'who spoke when' may not be enough. 
It is often more desirable to determine 'who spoke what'. This involves associating speaker ID labels with transcripts generated by an Automatic Speech Recognition (ASR) system, a process known as speaker-attributed ASR (SA-ASR). This typically requires a complex system architecture, with core modules for ASR and SD \cite{Kanda2020InvestigationOE, Kanda2021EndtoEndSA, Huang2023TowardsWE, Li2023SaParaformerNE}. Additionally, it is prevalent to include a speaker change detection (SCD) module before the SD module, which identifies the boundaries between different speakers turns and splits the ASR transcript into homogeneous token segments.

Stream-based SA-ASR is a challenging problem with limited prior research available.  Specifically, 
techniques like Turn-to-diarize \cite{Kanda2021TranscribetoDiarizeNS} use ASR-based speaker turn detection for segmentation, followed by multi-stage clustering \cite{Wang2022HighlyER} to further minimize memory usage in streaming applications.
Recently, streaming SA-ASR methods \cite{kanda22_interspeech, kanda22b_interspeech} based on the token-level serialized output training (t-SOT) strategy have demonstrated promising performance on simulated multi-talker datasets. However, in our preliminary studies, all of these methods exhibited reduced performance when applied to real spontaneous conversations, such as multi-speaker meetings.  
One key characteristic of spontaneous multi-speaker meetings is the significant variability in utterance lengths, ranging from a single word to several sentences, with some transitions lacking filled pauses between utterances. While SCD can create internally homogeneous segments from a speech stream, achieving accurate SD remains challenging. Short utterances often provide insufficient speaker information, leading to poor verification performance. This issue is exacerbated in hours-long streaming SA-ASR scenarios during meetings.

To tackle this practical challenge, we propose a streaming diarization system called SCDiar, which introduces a novel segment selection process that operates between the SCD and SD clustering stages. As demonstrated by our experimental results, this process significantly improves performance on the streaming SA-ASR task, particularly with real meeting data. This success can be attributed to three key innovations:
(1) the development of a length-aware similarity matrix, in contrast to the traditional cosine similarity matrix; 
(2) the introduction of a novel approach that, for the first time, selects only the best available segment (referred to as the representative segment) for each speaker, instead of using all available segments; and (3) the proposal of a mathematical optimization scheme to efficiently select these representative segments.
As shown in our experiments, with all these innovations, SCDiar outperforms state-of-the-art online SA-ASR systems and even achieves performance close to that of offline systems. 

\section{Our System}
\label{sec:system}

\begin{figure*}
    \centering
    \includegraphics[width=17.0cm]{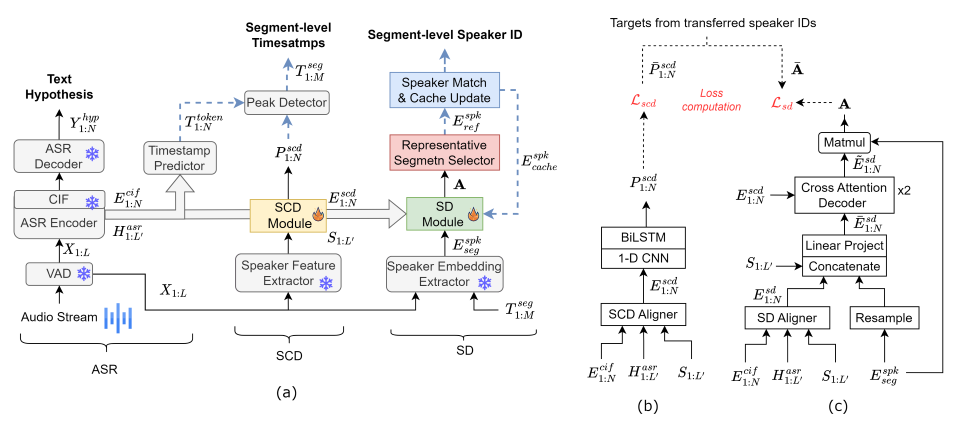}
    \caption{(a) The overview of inference process in the SCDiar system. (b) The structure of the SCD module. (c) The structure of the SD module.}
    \label{fig:system}
\end{figure*}

Figure \ref{fig:system}(a) illustrates the overview of our proposed SCDiar. This architecture consists of three main blocks: ASR, SCD and SD. Both SCD and SD modules, except for the modules related to SD inference, are trained together using a multi-target loss function.

\subsection{ASR}

A CIF-based \cite{cif_dong2020} ASR is employed as our backbone due to its powerful alignment capability between predicted hypothesis tokens and acoustic features. For a given speech stream, each speech chunk $X$, detected by VAD (subjecting to the maximum chunk length limitation $L_0$), is fed to the ASR encoder. Tw
Initially, the ASR encoder transforms the audio segment $X_{1:L}$ into a subsampled sequence of acoustic features $H^{\text{asr}}_{1:L'}$.
Following this, the CIF module estimates weights $\alpha$ for each frame in $H^{\text{asr}}_{1:L'}$ and produces the integrated embeddings ${E}^{\text{cif}}_{1:N}$. 
The ASR decoder then takes ${E}^{\text{cif}}_{1:N}$ as input and predicts hypothesis $Y^{\text{token}}_{1:N}$, consisting of $N$ tokens.
The process is summarized as follows:
\begin{align} 
    H^{\text{asr}}_{1:L'} &= \text{ASREncoder}(X_{1:L}) \label{eq:encoder}\\
    [\alpha_{1:L'}, {E}^{\text{cif}}_{1:N}] &= \text{CIF}(H^{\text{asr}}_{1:L'}) \label{eq:cif}\\ 
    Y^{\text{hyp}}_{1:N} &= \text{ASRDecoder}({E}^{\text{cif}}_{1:N}) \label{eq:decoder},
\end{align}
Based on the special mechanism of CIF, the alignment between $H^{\text{asr}}_{1:N}$ and ${E}^{\text{cif}}_{1:N}$ can be used to estimate the timestamps of tokens:
\begin{equation}
    T^{\text{token}}_{1:N} = \text{TimestampPredictor}(H^{\text{asr}}_{1:L'} {E}^{\text{cif}}_{1:N}) \label{eq:ts}
\end{equation}

\subsection{Token-level Speaker Change Detection}

Unlike SOT-based methods that employ ASR for outputting speaker turns, we design an auxiliary module for token-level SCD on a pre-trained ASR, which utilizes both prosody features (e.g. pause) and speaker-related features to enhance detection accuracy.
Initially, we use a speaker feature extractor to obtain a frame-level sequence $S_{1:L'}$. 
Then an aligner module containing a multi-head attention (MHA) layer is used to receive ${E}^{\text{cif}}_{1:N}$, $H^{\text{asr}}_{1:L'}$ and $S_{1:L'}$ as query, key and value inputs respectively, and output a token-level feature sequence $E^{\text{scd}}_{1:N}$, which is illustrated in Figure \ref{fig:system}(b).
Subsequently, a BiLSTM layer followed by an 1-D CNN layer with a kernal size of 3 refines this sequence. The probability sequence $P^\text{scd}_{1:N}$ for token-level SCD is determined using a softmax operation. 
Finally, peak detection\footnote{scipy.signal.find\_peaks} is applied to the sequence to detect the speaker change tokens with a threshold $\theta_{scd}$.
Along with token-level timestamps, we can obtain the timestamps $T^{\text{seg}}_{1:M}$ for the split $M$ segments.
\begin{align} 
    E^{\text{scd}}_{1:N} &= \text{SCDAligner}( {E}^{\text{cif}}_{1:N}, H^{\text{asr}}_{1:L'}, S_{1:L'}) \label{eq:scd}\\
    P^\text{scd}_{1:N} &= Softmax(\text{CNN}(\text{BiLSTM}(E^{\text{scd}}_{1:N}))) \\
    T^\text{seg}_{1:M} &= \text{PeakDetector}(P^\text{scd}_{1:N}, \theta_{scd}, T^{\text{token}}_{1:N})
\end{align}

\begin{figure}[h]
    \centering
    \includegraphics[width=8.5cm]{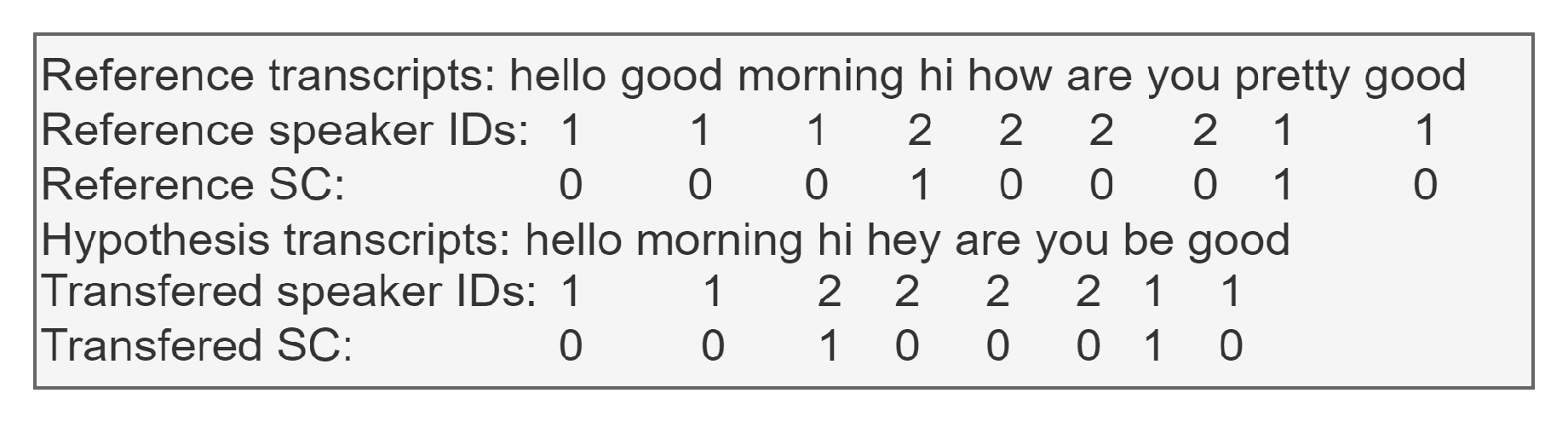}
    \caption{An example for TPST \cite{Wang2024DiarizationLMSD}.}
    \label{fig:tpst}
\end{figure}

During training, we freeze the parameters in ASR.
To alleviate the mismatch between the hypothesis and reference transcripts, we employ Transcript-Preserving Speaker Transfer (TPSP) algorithm \cite{Wang2024DiarizationLMSD} to obtain the target labels for SCD, which maps the reference speaker information to the hypothesis transcripts.
An example of the transferred speaker-attributed transcripts is shown in Figure \ref{fig:tpst}.
The focal loss function \cite{Lin_2017_ICCV} is applied to compute $\mathcal{L}_{scd}$ as SCD loss.

\subsection{Segment-token Similarity Matrix from SD Network}

Given $M$ segments from SCD, a common process for clustering is to compute the cosine similarity matrix.
However, the square similarity matrix disregards some important factors such as segment length, as segment length often correlates with the quality and richness of speaker information.
We instead compute the segment-token similarity matrix for clustering by considering the contribution of individual tokens within each segment.
We first employ a speaker embedding extractor to obtain segment-level embeddings $E^{\text{spk}}_{seg}$, i.e.,
\begin{align} 
    E^{\text{spk}}_{seg,1:M} = \text{EmbedExtractor}(X_{1:L}, T^{\text{seg}}_{1:M}), \label{eq:seg}
\end{align}
Then, we develop a network to refine the speaker similarity, as illustrated in Figure \ref{fig:system}(c).
Similar to Eq\eqref{eq:scd}, an aligner module is used to obtain token-level speaker embeddings $E^{\text{sd}}_{1:N}$.
With local information from $E^{\text{sd}}_{1:N}$ and global information from $E^{\text{spk}}_\text{seg}$, we apply a multi-scale approach to concatenate these features.
We resample the embeddings in $E^\text{spk}_\text{seg}$ according to the number of tokens in each segment and align them with $E^{\text{sd}}_{1:N}$ and $E^{\text{scd}}_{1:N}$.
A linear projection layer is then used for dimension reduction.
Then, two cross-attention decoder layers are used to generate the refined token-level speaker embeddings $\tilde{E}^{\text{sd}}_{1:N}$.
\begin{align} 
    E^{\text{sd}}_{1:N} &= \text{SDAligner}({E}^{\text{cif}}_{1:N}, H^{\text{asr}}_{1:L'}, S_{1:L'}) \label{eq:sd1}\\
    \bar{E}^{\text{sd}}_{1:N} &= \text{Project}([E^{\text{sd}}_{1:N}, E^{\text{scd}}_{1:N}, \text{Resample}(E^{\text{spk}}_{\text{seg}, 1:M})]) \label{eq:sd2}\\
    \tilde{E}^{\text{sd}}_{1:N} &= \text{CrossAttDecoder}(\bar{E}^{\text{sd}}_{1:N}, S_{1:L}) \label{eq:sd2}\\
                            &=\bar{E}^{\text{sd}}_{1:N} + \text{MHA}(\bar{E}^{\text{sd}}_{1:N}, S_{1:L}, S_{1:L}) \nonumber
\end{align}
Finally, the similarity matrix ${\bf{A}}\in\mathcal{R}^{N*M}$ is computed through matrix multiplication as follows:
\begin{align} 
   {\bf{A}} = Sigmoid((\tilde{E}^{\text{sd}}_{1:N})^T E^{\text{spk}}_{\text{seg}, 1:M}), \label{eq:mat}
\end{align}
where ${\bf{A}}_{i,j}$ denote the similarity score between the speakers associated with the $i$-th token and the $j$-th segment.
An example is depicted in Figure 3, where the refined matrix {\bf{A}} are more distinct compared to the matrix derived from cosine distance. 
Importantly, the segment length information is preserved in this rectangle matrix, which is crucial for the subsequent quality evaluation and selection process.


During training, we use reference SCD results for segmentation in Eq\eqref{eq:seg}.
A weighted binary cross-entropy (BCE) loss is computed between the predicted matrix {\bf{A}} and target binary matrix ${\bf{\bar{A}}}$ while taking into account the token count $N_j$ within the $j$-th segment.
\begin{align} 
   \mathcal{L}_{sd} = \sum_{j=1}^M (\frac{N_j}{N}\text{BCE}({\bf{A}}[:,j], \bar{\bf{A}}[:,j])) \label{eq:mat}
\end{align}

\begin{figure}
    \centering
    \includegraphics[width=9.0cm]{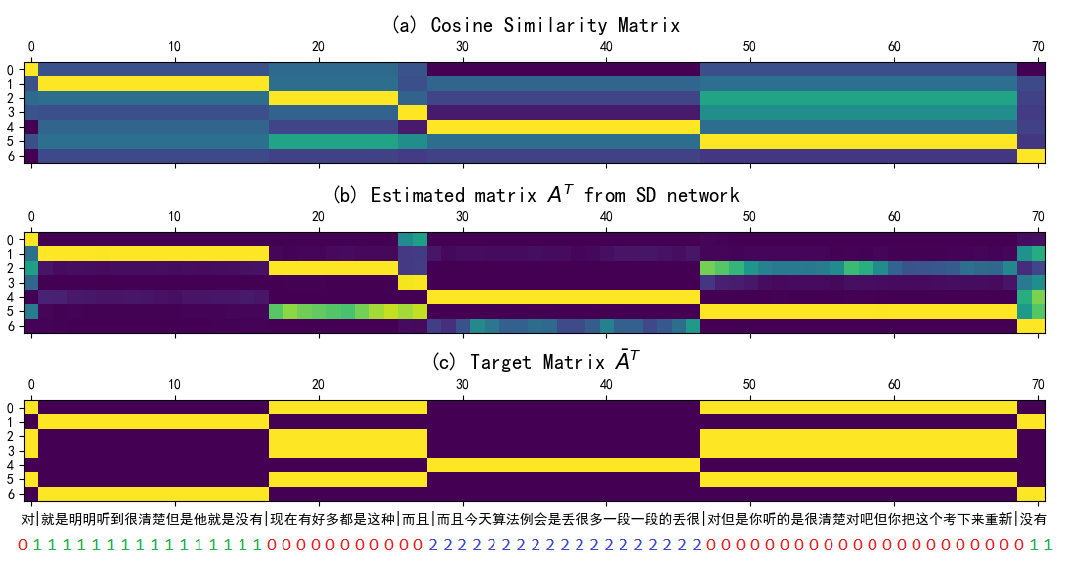}
    \caption{ Segment-token similarity matrix from (a) cosine distance, (b) estimated ${\bf{A}}^T$ and (c) target ${\bar{\bf{A}}}^T$. Reference speaker IDs are listed at the bottom with 7 segments.}
    \label{fig:matrix}
\end{figure}

\subsection{Representative Segments Selection via Optimization}

To leverage the rich information in matrix A, we design a novel selection approach aimed at identifying
the \textit{``representative segment"} for each speaker. 
This approach convert the many-to-one mapping into an one-to-one mapping problem, which is helpful to cluster insufficient data points in streaming mode.
The \textit{representative segment} serves as an anchor point for clustering, which should be stable (a long enough duration), be close to segments from the same speaker, and remain distinct from segments of other speakers.
To achieve this, we design an optimization problem for selecting \textit{representative segments} from $M$ segments, which is formulated as:
\begin{align}
    \text{arg}\min\limits_x  \|{\bf{A}}x -\mathbf{1}\|_2^2, \ \ \text{s.t}\ \ \ x_i \in \{0, 1\} \label{eq:prog},
\end{align}
where a binary vector $x\in\mathcal{Z}^{M}$ denotes the distribution of \textit{representative segments} and $\mathbf{1}\in\mathcal{Z}^{N}$ is an all-one vector. 
The number of ones in $x$ equals to the predicted number of speakers.
Since we want each token assigned to exactly one speaker, the objective function in Eq\eqref{eq:prog} penalizes the speaker overlap (${\bf{A}}{[i,:]}x > 1$) and miss error (${\bf{A}}{[i,:]}x < 1$) for the $i$-th tokens.

Given that the integer constraints in Eq\eqref{eq:prog} can be computationally intensive when $M$ is large, we relax Eq\eqref{eq:prog} into a non-negative least squares programming with boundary constraints \cite{Stark2008BoundedVariableLA}, which can be efficiently solved\footnote{http://docs.scipy.org/doc/scipy}:
\begin{align}
    \text{arg}\min\limits_x  \|{\bf{A}}x -\mathbf{1}\|_2^2, \ \ \text{s.t}\ \ \ 0 \leq x_i \leq 1 \label{eq:prog2}.
\end{align}
A threshold $\theta_{sd}$ is then applied to $x$ to select the \textit{representative segments}, with their embeddings denoted as $E^{spk}_{rep}$.


\subsection{Speaker Label Mapping and Cache Update}
\label{ssec:cache}

To track the speaker statistics in a streaming fashion, we use a cache (buffer) to store the embedding centers of existing speakers, which are denoted as $E^{spk}_\text{cache}$.
Taking this into account, we modify Eq\eqref{eq:mat} and Eq\eqref{eq:prog2} as follows:
\begin{align} 
   {\bf{A}} &= Sigmoid((\tilde{E}^{\text{sd}}_{1:N})^T [E^{spk}_{1:M}; E^{spk}_\text{cache}]), \label{eq:mat2} \\
    \text{arg}\min\limits_{x, x_{c}} &\|{\bf{A}}[x^T;x_{c}^T]^T -\mathbf{1}\|_2^2, \ \ \text{s.t}\ \ \ 0 \leq x_i,x_{c,i} \leq 1 \label{eq:program3},
\end{align}
where $x_{c}$ refers to the \textit{representative segments} selected for cached speakers.

The final mapping and update process involves several steps:
1) Calculate cosine distances between $E^\text{spk}_\text{rep}$ and $E^\text{spk}_\text{cache}$ and discard embeddings in $E^\text{spk}_\text{rep}$ that are highly similar to cached ones using a threshold $\theta_\text{cache}$;
2) Update $E^\text{spk}_\text{cache}$ by appending the remaining $E^\text{spk}_\text{rep}$ as new speakers;
3) Map speakers IDs to the $M$ segments by comparing $E^\text{spk}_\text{cache}$ and $E^\text{spk}_\text{seg}$ using cosine distance;
4) Update the embedding center $E^\text{spk}_\text{cache}$ with the matched embeddings among $E^\text{spk}_\text{seg}$, weighting them by token count, i.e.,
\begin{align*}
    E^{spk}_{cache,i} & \leftarrow  (N_{c,i} E^{spk}_{cache,i} + N_m E^{spk}_{seg,m})/(N_{c,i} + N_{m}) \\
    N_{c,i} & \leftarrow  N_{c,i} + N_{m},
\end{align*}
where $N_m$ is the token count of matched segment in the current stream for the $i$-th cache speaker and $N_{c,i}$ is the total token count of the matched segments in the previous streams.

\subsection{Training Strategy}
\label{ssec:training}

During training, a multi-target function is applied to optimize the SCD and SD modules.
Since focal loss for SCD usually has small values, we use a factor $\lambda$ for scaling.
\begin{equation}
    \mathcal{L} = \lambda\cdot\mathcal{L}_{scd} + \mathcal{L}_{sd} \label{eq:loss}.
\end{equation}

\noindent\textbf{Split strategy:} To simulates the variant possible results from SCD in Eq\eqref{eq:seg}, an augmentation strategy is applied to randomly divide a long segment in a sample into two short segments during training. 

\section{Experiments}
\label{sec:experiment}

\subsection{Dataset}
For training, we use the 2-speaker conversation corpus \textit{ASR-BIGCCSC} which comprises over 5,000 hours of Mandarin speech.\footnote{https://magichub.com/datasets/mandarin-chinese-conversational-speech-corpus-spontaneous-conversation}
We divide long recordings into segments of 4$\sim$30 seconds based on sentences-level annotations. To augment the variety of speakers in the samples, we randomly concatenate two segments from different conversations via an on-the-fly mode. 

For evaluation, the test set of AISHELL-4 corpus \cite{fu21b_interspeech} is used, which consists of 20 recorded meeting sessions with 4 to 8 speakers per session, and the total length is 12.7 hours. The BeamformIt \cite{Anguera2007} technique was employed to transform multi-channel signals into single-channel signals.
Additionally, we gathered a more challenging in-house meeting corpus for evaluation, which consists of 5 internal real meetings spanning a total of 7.0 hours. Each meeting lasts at least an hours and involves at least 10 speakers.

\subsection{Experimental Setup}

An online VAD\footnote{iic/speech\_fsmn\_vad\_zh-cn-16k-common-pytorch} is employed before ASR, where the maximum active segment length is set to 15 seconds.
The pre-trained Paraformer model\footnote{iic/speech\_paraformer-large\_asr\_nat-zh-cn-16k-common-vocab8404-pytorch} \cite{gao22b_interspeech}  is used as the CIF-ASR backbone. 
The CAM++ model\footnote{iic/speech\_campplus\_sv\_zh\_en\_16k-common\_advanced} \cite{wang23ha_interspeech} is used to extract speaker features $S_{1:L'}$ in Eq\eqref{eq:scd} and speaker embeddings $E^{spk}_{1:M}$ in Eq\eqref{eq:seg}, where $S_{1:L'}$ are obtained from the layer before the statistics pooling layer.
The dimension of attention and feed-forward layers are set to 256 and 2048, respectively.
The multi-head attention layers in Eq\eqref{eq:scd} and \eqref{eq:sd1} have 4 heads with 512 attention dimension and 1024 feed-forward dimensions.
While in Eq\eqref{eq:sd2} the dimensions change to 192, 256 to match the dimension of speaker embeddings.
The hidden size in the BiLSTM is also 192.

During training, the $\alpha$ and $\gamma$ in the focal loss function $\mathcal{L}_{scd}$ are set to 0.25 and 2 respectively, and $\lambda$ in Eq\eqref{eq:loss} is set to 10. The network are trained for at most 15 epochs, and the last 5 modules are averaged for evaluation.
During inference, $\theta_{scd}$ in Eq\eqref{eq:scd} is set to 0.25, and $\theta_{sd}$ in Eq\eqref{eq:prog2} is set to 0.3 experimentally. In Section\ref{ssec:cache}, the threshold $\theta_\text{cache}$ in step1 is set to 0.55, and the segments containing fewer than 10 tokens will be excluded for representative segment selection and cache update.

To evaluate the diarization performance, two common speaker attributed metrics are used: the Word Diarization Error Rate (WDER) \cite{shafey19_interspeech} and the concatenated minimum-permutation word error rate (cpWER) \cite{Watanabe2020CHiME6CT}.

\subsection{Results}
\label{ssec:res}

\begin{table}[htbp]
  \centering
  \caption{Results over two test sets. WERs on AISHELL-4 and In-house datasets are 18.45\% and 17.40\% respectively. SpC: spectral clustering. $\Delta$cpWER: cpWER-WER}
  \scalebox{0.95}{
    \begin{tabular}{r|l|cc|cc}
    \hline
    No      &  Method  & \multicolumn{2}{c|}{AISHELL-4} & \multicolumn{2}{c}{In-house} \\
    \cline{3-6}    &    & $\Delta$cpWER & WDER  & $\Delta$cpWER & WDER \\
    \hline
    \multicolumn{2}{l|}{\textit{Offline methods}} &       &       &       &  \\
    1     & Sliding Window+SpC & 2.13  & 2.69  & 9.95  & 13.51  \\
    2     & SCD Segment+SpC & 2.13  & 2.87  & 11.32  & 14.75 \\
    \hline
    \multicolumn{2}{l|}{\textit{Online methods}} &       &       &       &  \\
    3     & Multi-stage (org.) & 48.31  & 34.09  & 75.98  & 51.30  \\
    4     & Multi-stage (mod.) & 27.75  & 18.40  & 64.29  & 47.20  \\
    5     & Core sample+VBx & 10.09  & 7.84  & 21.65  & 19.23  \\
    6     & SCDiar (proposed) & \textbf{3.42}  & \textbf{3.56}  & \textbf{10.66}  & \textbf{15.36}  \\
          & \quad -w.o. split strategy & 3.43  & 3.59  & 14.38  & 16.89  \\
          & \quad -w.o. rep. select & 11.96  & 8.76  & 13.63  & 16.29  \\
    \hline
    \hline
    \end{tabular}%
    }
  \label{tab:res}%
\end{table}%

Table \ref{tab:res} compares the performance on different methods using the same ASR and speaker embedding extractor.
Among the offline methods, System 1 uses sliding window of 1.5s with 0.25s jump for segmentation to extract speaker embeddings followed by spectral clustering. 
The predicted speakers labels are subsequently aligned to the segments based on the timestamps.
While System 2 uses our SCD results for segmentation, and the segments shorter than 1 second are filtered before spectral clustering for better performance.
We also reproduced two online clustering approaches for comparison: Multi-stage Clustering \cite{Huang2023TowardsWE} and core-samples selection with VBx \cite{Yue2022OnlineSD}.
For Multi-stage Clustering methods, we used our SCD module for speaker turn detection then combined it with the multi-stage clustering strategy from the original implementation\footnote{https://github.com/wq2012/SpectralCluster}.
To be fair, we also replaced the main clusterer in System 3 by the spectral clusterer used in System 1,2 and named it as System 4.
As for core samples selection with VBx, the block size was set to 120 as that in the paper, and the PLDA parameters in VBx were adopted to our speaker embeddings extractor based on the Voxceleb-2 dataset \cite{chung2018voxceleb2}.

From the table, we can obtain a few valuable insights:
1) The number of speakers in the meetings has a huge impact on the diarization results, where cpWER on In-house (10+ speakers) are much larger than that on AISHELL-4 (4$\sim$8 speakers);
2) Although System 2 using SCD results for segmentation performs slight worse than System 1, the SCD results with length information can be really helpful for online systems. 
Our method with representative segment selection surpasses the other online approaches and significantly narrows down the gap between online methods and offline methods with only a marginal 1.3\% and 0.7\% increase in $\Delta$cpWER.
3) In the ablation study, we found that split strategy in Section \ref{ssec:training} can improve the performance which benefits similarity computation in SD.
4) Without representative segments, the performance drops significantly on both datasets, which proves the essentiality of the proposed SD network and optimization process.

\begin{figure}
    \centering
    \includegraphics[width=8.5cm]{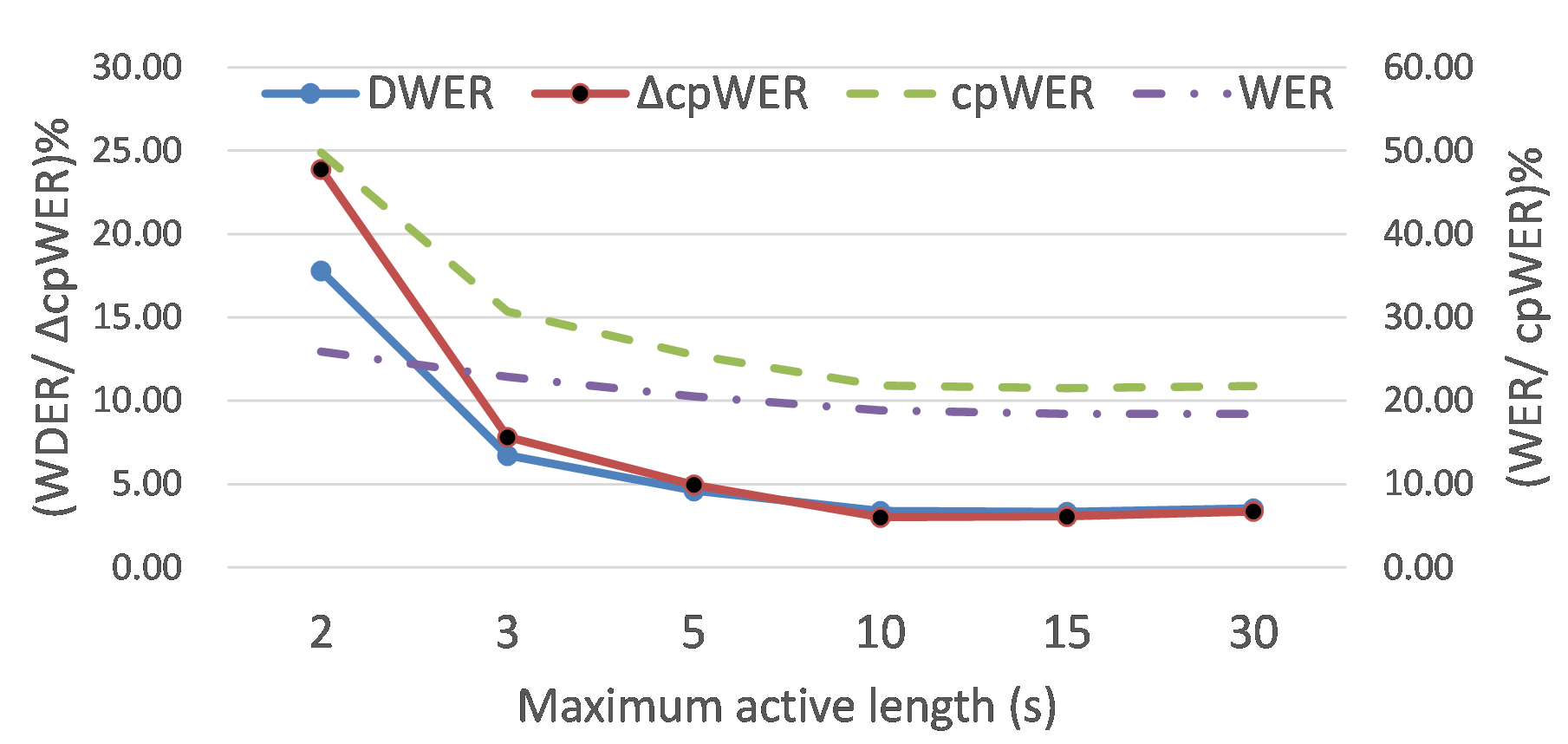}
    \caption{ASR and diarization results on AISHELL-4 with different maximum active segment lengths of the VAD.}
    \label{fig:curve}
\end{figure}

Since the maximum latency of the whole system is determined by the maximum active segment length of VAD, we also evaluate the performance using different VAD settings, as shown in Figure \ref{fig:curve}.
When the streaming input is shorter than 3 seconds, the number of the segments and their short duration may not be sufficient for representative segment selection, which leads to large WDER and cpWER. 
However, when the length is larger than 3 seconds, the performance curve fast converges and become almost horizontal after 10 seconds.

To show the computation complexity of our method, we measure real-time factor (RTF) using inference on an Intel(R) Xeon(R) Gold 6148 CPUs @ 2.40GHz and an NVIDIA V100.
The average values on the in-house data for the ASR, SCD and SD module are about 0.072, 0.004 and 0.009 respectively, which satisfies real-time processing. 

\section{Conclusion}
\label{sec:conclusion}

This paper proposes a streaming system that integrates both ASR and diarization.
With manageable inference latency and computational cost,
our method approaches the performance of offline systems in hours-long meetings with over 10 speakers.
Several important issues remain unaddressed in our system, such as handling speech overlaps, variations in voice caused by speaker emotions, spatial positions, or head movement, which can result in the creation of erroneous 'virtual' speakers. We will tackle these challenges in future work.





\bibliographystyle{IEEEtran}
\bibliography{refs.bib}

\end{document}